\documentclass[11pt,onecolumn,amssymb,nofootinbib]{revtex4}
\usepackage{amsmath, amsthm, amscd, amssymb}
\usepackage{bm}
\usepackage{bbm}

\begin{document}
\title{\bf Adler sum rule}
\author{Stephen L. Adler}
\email{adler@ias.edu} \affiliation{Institute for Advanced Study,
Einstein Drive, Princeton, NJ 08540, USA}

\begin{abstract}
The {\bf Adler sum rule} states that the integral over energy of a difference of
neutrino-nucleon and antineutrino-nucleon structure functions is a constant, independent
of the four-momentum transfer squared.  This constancy is a  consequence of the
local commutation relations of the time components of the hadronic weak current, which follow
from the underlying quark structure of the standard model.

\end{abstract}

\maketitle
\bigskip
\leftline{\bf Contents}

\leftline{1. Statement of the Adler sum rule}
\leftline{2. Relation to the Adler-Weisberger and Cabibbo-Radicati sum rules,}
\leftline{~~~and the Bjorken electron scattering inequality }
\leftline{3. Saturation of the sum rule and Bjorken scaling}
\leftline{4. Sketch of derivation}

\section{Statement of the Adler sum rule}

Consider the inclusive neutrino-nucleon or antineutrino-nucleon scattering reactions
\begin{equation}\label{eq:scattering}
\nu/\bar\nu(k) + N(p) \to \ell^{-/+}(k') +X(p')~~~,
\end{equation}
with $\ell^{-/+}$ the lepton corresponding to the incident
neutrino/antineutrino, and with $X$ an unobserved  hadronic final
state. Since the lepton in cases of greatest interest is an electron
or muon, the lepton mass can be neglected. Defining the
four-momentum transfer and energy transfer variables $q$ and $\nu$
by
\begin{equation}\label{eq:qandnu}
q=k-k'~,~~~\nu=-p\cdot q/M_N~~~,
\end{equation}
with $M_N$ the nucleon mass, one finds in the laboratory frame where
the initial nucleon is at rest, using a $(+,+,+,-)$ metric convention,
\begin{align}\label{eq:labframe}
p=&(M_N,\vec 0)~,~~~k=(E,\vec k)~,~~~k'=(E',\vec k')~~~,\cr
\nu=&E-E'~,~~~q^2=4EE'\sin^2(\theta/2) ~~~,\cr
\end{align}
with $\theta$ the angle between $\vec k'$ and $\vec k$.  Analysis of
the kinematic structure of the reaction of Eq.~\eqref{eq:scattering}
shows that the inclusive cross section $d^2 \sigma/d(q^2)d\nu$ takes
the form
\begin{equation}\label{eq:crosssection}
\frac {d^2 \sigma^{\nu/\bar\nu}  } {d(q^2)d\nu}=\frac{G_F^{\,2}}{2\pi}\frac{E'}{E}
\left[2 W_1^{\nu/\bar\nu}(q^2,\nu) \sin^2(\theta/2)+W_2^{\nu/\bar\nu}(q^2,\nu) \cos^2(\theta/2)
+\epsilon^{\nu/\bar\nu} W_3^{\nu/\bar\nu}(q^2,\nu) \frac{E+E'}{M_N}  \sin^2(\theta/2) \right]~~~,
\end{equation}
with $\epsilon^{\nu/\bar\nu}=-1/+1$, with $G_F$ the Fermi weak
interaction constant (assuming that $q^2$ is much smaller than the
charged intermediate boson mass squared), and with $W_{1,2,3}$ the
{\bf structure functions} for deep inelastic neutrino scattering. In
terms of the $W_2$ structure function, the Adler sum rule
\cite{adler1} takes the form
\begin{equation}\label{eq:sumrule1}
K_N=\int_0^{\infty} d\nu [W_2^{\bar \nu}(q^2, \nu) - W_2^{\nu}(q^2,\nu)]~~~,
\end{equation}
with $K_N$ a constant. (The lower integration limit can be taken as just below
the single nucleon contribution at $\nu=q^2/(2M_N)$, instead of 0.)  For a proton target  $K_{N={\rm proton}}=2$,
while for a neutron target $K_{N={\rm neutron}}=-2$.  When
production of heavy flavors such as charm is neglected, the
corresponding expressions for $K_N$ are $K_{N={\rm proton}}=2
 +2 \sin^2 \theta_C$ and $K_{N={\rm neutron}}=-2
 + 4 \sin^2 \theta_C$, with $\theta_C$ the Cabibbo
angle; since the Adler sum rule was derived well before the
discovery of charm, some older texts give these expressions for
$K_N$. In his original paper \cite{adler1}, Adler used a different
notation from the one now standard, labelling $W_1$ as $\alpha$,
$W_2$ as $\beta$ and $W_3$ as $2M_N\gamma$, multiplied by the appropriate
Cabibbo angle factors $\cos^2 \theta_C$ and $\sin^2\theta_C$ in the
strangeness conserving and strangeness changing cases, respectively,
which were treated separately.

According to Eq.~\eqref{eq:labframe}, as the neutrino energy
approaches infinity, for fixed $q^2$ one has $\sin^2(\theta/2) \to
0$ and $\cos^2(\theta/2) \to 1$.  Hence in this limit the deep
inelastic cross section is dominated by the $W_2$ structure
function, and so integrating over the energy transfer $\nu$,
Eq.~\eqref{eq:sumrule1} yields the limiting relation
\begin{equation}\label{eq:limit1}
\lim_{E_{\nu}\to \infty} \left[ \frac {d\sigma^{\bar  \nu p}}{d(q^2)}-\frac {d\sigma^{  \nu p}}{d(q^2)}\right]=\frac{G_F^2}{\pi}~~~,
\end{equation}
and similarly (with a reversal of sign) for the difference of
antineutrino and neutrino differential cross sections on a neutron
target.
\section{Relation to the Adler-Weisberger and Cabibbo-Radicati sum rules, and the Bjorken electron scattering inequality}

Equation \eqref{eq:sumrule1} is the sum of axial-vector and vector
sum rules, which can be written separately in terms of the
corresponding contributions to the structure function $W_2$, denoted
in what follows by the subscripts $V, A$ respectively. Neglecting heavy flavor production and approximating
$\sin^2 \theta_C\simeq 0$, the
axial-vector part of Eq.~\eqref{eq:sumrule1}, on a proton target, is
\begin{equation}\label{eq:sumrule2}
1=g_A(q^2)^2+\int_{\nu_{\rm th}}^{\infty} d\nu [W_{2A}^{\bar \nu}(q^2, \nu) - W_{2A}^{\nu}(q^2,\nu)]~~~,
\end{equation}
while the vector part of Eq.~\eqref{eq:sumrule1} is
\begin{equation}\label{eq:sumrule3}
1=F_{1V}(q^2)^2+q^2F_{2V}(q^2)^2+\int_{\nu_{\rm th}}^{\infty} d\nu [W_{2V}^{\bar \nu}(q^2, \nu) - W_{2V}^{\nu}(q^2,\nu)]~~~.
\end{equation}
Here $\nu_{\rm th}=(M_{\pi}^2+2M_NM_{\pi}+q^2)/(2M_N)$ denotes the
pion-nucleon continuum threshold, and the nucleon contributions
have been explicitly separated off in terms of the nucleon
axial-vector form factor $g_A(q^2)$ and the nucleon isovector
electromagnetic form factors $F_{1V}(q^2)$ and $F_{2V}(q^2)$.

At $q^2=0$, the axial-vector sum rule of Eq.~\eqref{eq:sumrule2}
becomes
\begin{equation}\label{eq:sumrule4}
1=g_A(0)^2+\int_{\nu_{\rm th}}^{\infty} d\nu [W_{2A}^{\bar \nu}(0, \nu) - W_{2A}^{\nu}(0,\nu)]~~~.
\end{equation}
According to the Adler forward lepton theorem \cite{adler2},
neutrino reactions with a forward-going lepton, in the approximation
of neglecting the lepton mass, can be expressed in terms of
corresponding pion reaction cross sections for zero mass pions. Thus
the  integrand of Eq. ~\eqref{eq:sumrule4} can be written in terms
of pion proton scattering cross sections as
\begin{equation}\label{eq:forward}
[W_{2A}^{\bar \nu}(0, \nu) - W_{2A}^{\nu}(0,\nu)]=\frac{2M_N^2g_A(0)^2}{\pi g_r(0)^2\nu}[\sigma^{\pi^-p}(0,\nu)-\sigma^{\pi^+p}(0,\nu)]
~~~,
\end{equation}
with $g_r(0)$ the off-shell pion-nucleon coupling constant.
Substituting this into Eq.~\eqref{eq:sumrule4} gives \cite{adler3}
the off-shell version of the earlier Adler \cite{adler4}-Weisberger
\cite{weis} sum rule, which is a consequence of the spatially integrated axial
charge current algebra.   The  on-shell
Adler-Weisberger sum rule, which is obtained \cite{adler4}, \cite
{weis} by extrapolating to physical mass pions using the partially
conserved axial-vector current (PCAC) hypothesis, gives a sum rule
for the axial vector coupling $g_A(0)$ that agrees well with
experiment.

Because the isovector vector charge is
conserved when the small up and down quark masses are neglected,
the continuum contribution to the vector sum rule of Eq.~\eqref{eq:sumrule3} vanishes at
$q^2=0$, where this sum rule reduces to the trivial identity $1=1$.
However, the first derivative of this sum rule at $q^2=0$ gives the
interesting  Cabibbo-Radicati \cite{cabibbo} sum rule,
\begin{equation}\label{eq:sumrule5}
0=2\frac{d}{d(q^2)}F_{1V}(q^2)|_{q^2=0}+F_{2V}(0)^2 +\int_{\nu_{\rm th}}^{\infty} d\nu
 \frac {d}{d(q^2)}[W_{2V}^{\bar \nu}(q^2, \nu) - W_{2V}^{\nu}(q^2,\nu)]|_{q^2=0}~~~.
\end{equation}

With application to the Stanford Linear Accelerator Center (SLAC)
electron scattering experiments in mind, Bjorken \cite{bj1}
converted the limiting relation Eq.~\eqref{eq:limit1} to a limiting
inequality for electron scattering.  This is possible because, since
the neutrino scattering cross section is positive,
Eq.~\eqref{eq:limit1} gives a lower bound for the antineutrino
proton scattering cross section, which implies a factor of 2 smaller
lower bound for the vector current contribution alone.   Since the
vector weak current is related by an isotopic spin rotation to the
isovector part of the electromagnetic current, Bjorken was then able
to obtain a lower bound to the sum of electron scattering cross
sections on a proton and a neutron, since in this sum the isovector
and isoscalar currents contribute incoherently.   Keeping track of
coupling constant and photon propagator factors, the resulting
electron scattering limiting inequality reads
\begin{equation}\label{eq:limit2}
\lim_{E_{e}\to \infty} \left[ \frac {d\sigma^{e p}}{d(q^2)}+\frac {d\sigma^{e n}}{d(q^2)}\right] > \frac{2 \pi \alpha^2}{(q^2)^2}~~~,
\end{equation}
with $\alpha \simeq 1/137$ the fine structure constant.

\section{Saturation of the sum rule and Bjorken scaling}

The salient feature of the sum rule of Eqs.~\eqref{eq:sumrule1},
\eqref{eq:sumrule2}, \eqref{eq:sumrule3}, is that the integral over
energy of the cross sections on the right gives a constant that is
independent of the momentum transfer squared $q^2$.  This is a very
different behavior from that of the nucleon  contributions, which
involve form factors that decrease rapidly to zero as $q^2$ is
increased. Moreover,  the low lying pion-nucleon resonance
contributions to the right hand side are known to have a
large $q^2$  behavior similar to that of the nucleon contributions. Thus, it was clear from
early on that a qualitatively new behavior would be needed for
saturation of the sum rule. Since structureless  particles have form
factors of unity rather than rapidly decreasing form factors, early
discussions also suggested that saturation of the sum rule would
indicate the existence of elementary constituents within the
nucleon.

The precise mechanism by which the sum rules are saturated was
clarified by the proposal by Bjorken \cite{bj2} of the Bjorken
scaling hypothesis, which states that in the limit of large $q^2$
and $\nu$, with $q^2/\nu$ fixed,  the structure functions
$W_1(q^2,\nu)$ and $W_2(Q^2,\nu)$ become functions of a single
scaling variable $x=q^2/(2M_N\nu)$, according to
\begin{equation}\label{eq:bjscaling}
\nu W_2(q^2,\nu)\to F_2(x)~,~~~M_N W_1(q^2,\nu)\to F_1(x)~~~.
\end{equation}
Since $d\nu=-\nu dx/x$, while $x=0$ at $\nu=\infty$, and $x=1$ at
threshold $\nu_{\rm th}$ in the scaling limit, in this limit the sum
rule of Eq.~\eqref{eq:sumrule1} becomes
\begin{equation}\label{eq:sumrulebj}
K_N=\int_0^1(dx/x) [F_2^{\bar \nu}(x) - F_2^{\nu}(x)]~~~,
\end{equation}
and the $q^2$-independence becomes manifest. Thus, saturation of the
sum rule requires  contributions from ever higher energies $\nu$ as
$q^2$ is increased to large values.   As discussed in the article on
Bjorken scaling, scaling is verified experimentally in deep
inelastic neutrino and electron scattering, up to small logarithmic
corrections, and was an important precursor of both the parton model
and quantum chromodynamics, in which the nucleon is a composite
constructed from point-like quark constituents. The Adler sum rule,
which is an exact relation even when scaling violations are taken
into account, has  been tested and verified experimentally,
providing direct evidence for the validity of the Gell-Mann
\cite{murray} local current commutator algebra of the weak hadronic
currents, which is the basis for the construction of the Yang-Mills
electroweak theory.

\section{Sketch of derivation}

To derive the sum rule of Eq.~\eqref{eq:sumrule1}, start from the expression
\begin{equation}\label{eq:comm1}
(2\pi)^{-1}\int dq_0 \int d^4 x e^{i q\cdot x}{\bar \Sigma}_s \langle N(p),s|[J_{h0}(x),J_{h0}^{\dagger}(0)] |N(p),s\rangle~~~,
\end{equation}
with $|N(p),s\rangle$ the state of a nucleon with four-momentum $p$ and spin $s$, and
with ${\bar \Sigma}_s$ denoting the spin average $(1/2)\sum_s$.
Here $J_{h0}(x)$ is the time component of the hadronic weak current, which is given by
\begin{equation}\label{eq:current1}
J_{h0}(x)=\sum_{U=u,c,t}\sum_{D=d,s,b} U^{\dagger}(x)(1-\gamma_5)V_{UD} D(x)~~~,
\end{equation}
with $V_{UD}$ elements of the Cabibbo-Kobayashi-Maskawa (CKM) flavor mixing matrix.
The commutator in Eq.~\eqref{eq:comm1} contains three types of terms, containing either no factors of
$\gamma_5$, one factor of $\gamma_5$, or two factors of $\gamma_5$.  Since $(\gamma_5)^2=1$, the
terms with two factors of $\gamma_5$  make a contribution
equal to the terms with no factors of $\gamma_5$, while the terms with one factor of $\gamma_5$
vanish after averaging over the spin $s$.  Thus Eq.~\eqref{eq:comm1} simpifies to
\begin{equation}\label{eq:comm2}
2 (2\pi)^{-1}\int dq_0 \int d^4 x e^{i q\cdot x} {\bar \Sigma}_s\langle N(p),s|[J_{h0}^{V}(x),J_{h0}^{V\dagger}(0)] |N(p),s\rangle~~~,
\end{equation}
with $J_{h0}^V(x)$ the time component of the vector part of the hadronic weak current,
given by
\begin{equation}\label{eq:current2}
J_{h0}^V(x)=\sum_{U=u,c,t}\sum_{D=d,s,b} U^{\dagger}(x)V_{UD} D(x)~~~.
\end{equation}
Since $(2\pi)^{-1}\int dq_0 e^{-iq_0x_0} =\delta(x_0)$, Eq.~\eqref{eq:comm2} involves only
equal time commutators, which can be evaluated by the fermion field canonical anti-commutation relations.
Dropping flavor off-diagonal contributions, which vanish when sandwiched between nucleon states,
the only commutator needed is
\begin{equation}\label{eq:commexample}
[U^{\dagger}(\vec x,0)D(\vec x,0),D^{\dagger}(\vec 0,0)U(\vec 0,0)]=\delta^3(\vec x) [U^{\dagger}(0)U(0)
-D^{\dagger}(0)D(0)  ]~~~.
\end{equation}
The appearance in this commutator of  $\delta^3(\vec x)$   eliminates the spatial integration in Eq.~\eqref{eq:comm2}, so
what remains is
\begin{equation}\label{eq:comm3}
2 {\bar \Sigma}_s \langle N(p),s|{\cal M} |N(p),s\rangle~~~.
\end{equation}
Here ${\cal M}$ is a linear combination of quark number operators, denoted by $n$ with the appropriate subscript,
multiplied by absolute value squared CKM matrix elements,
\begin{align}\label{eq:comm4}
{\cal M}=&\sum_{U=u,c,t}\sum_{D=d,s,b} |V_{UD}|^2(n_U-n_D) \cr
=&|V_{ud}|^2(n_u-n_d)+|V_{us}|^2(n_u-n_s)+|V_{ub}|^2(n_u-n_b)\cr
+&|V_{cd}|^2(n_c-n_d)+|V_{cs}|^2(n_c-n_s)+|V_{cb}|^2(n_c-n_b)\cr
+&|V_{td}|^2(n_t-n_d)+|V_{ts}|^2(n_t-n_s)+|V_{tb}|^2(n_t-n_b)~~~.\cr
\end{align}
Since a proton contains $n_u=2$ up quarks and $n_d=1$ down quark, and
a neutron contains $n_u=1$ down quark and $n_u=2$ up quarks, with  zero
quark number for $s, c, b, t$ type quarks, substituting Eq.~\eqref{eq:comm4}
into Eq.~\eqref{eq:comm3} gives for $N={\rm proton}$,
\begin{equation} \label{eq:semifinal1}
K_{N={\rm proton}}\equiv 2{\bar \Sigma}_s \langle N(p),s|{\cal M} |N(p),s\rangle = 2 [|V_{ud}|^2 +2(|V_{us}|^2+|V_{ub}|^2)-(|V_{cd}|^2+|V_{td})|^2)]~~~,
\end{equation}
and gives for $N={\rm neutron}$,
\begin{equation} \label{eq:semifinal2}
K_{N={\rm neutron}}\equiv 2{\bar \Sigma}_s \langle N(p),s|{\cal M} |N(p),s\rangle = 2[- |V_{ud}|^2 +|V_{us}|^2+|V_{ub}|^2-2(|V_{cd}|^2+|V_{td})|^2)]~~~.
\end{equation}
Finally, substituting the unitarity relations for the CKM matrix elements,
\begin{align}\label{eq:unitarity}
|V_{ud}|^2|+|V_{us}|^2+|V_{ub}|^2=&1~~~,\cr
|V_{ud}|^2+|V_{cd}|^2+|V_{td}|^2=&1~~~,\cr
\end{align}
Eqs.~\eqref{eq:semifinal1} and \eqref{eq:semifinal2} reduce to
$K_{N={\rm proton}}=2$ and $K_{N={\rm neutron}}=-2$.

The remainder of the derivation consists of relating Eq.~\eqref{eq:comm1} to an integral
over a difference of  neutrino and antineutrino scattering structure functions. In \cite{adler1} this was done
by working in the nucleon rest frame ($\vec p=0$) and postulating an unsubtracted dispersion
relation, which is valid for the $\beta = W_2$ sum rule case.   In the more recent textbook
and review article treatments referenced below, this is done by taking the limit of an infinite momentum ($|\vec p| \to \infty$) frame inside the $q_0$ integral, which uniquely picks out the $W_2$ structure function contribution.  Both methods
give the result quoted in Eq.~\eqref{eq:sumrule1}.  Both \cite{adler1} and the infinite momentum frame derivations  referenced
below omit heavy quark flavors and use the  Gell-Mann $SU(3)$
current algebra to evaluate the hadronic current commutators, rather than the full CKM matrix formulation used
here.

\section{Acknowledgement}
   The work of S.L.A. was
supported by the Department of Energy under grant no
DE-FG02-90ER40542.

\end{document}